%% file: paper.tex
\documentclass[conference]{IEEEtran}

\usepackage{cite}
\usepackage{amsmath,amssymb,amsfonts}
\usepackage{graphicx}
\usepackage{multirow}
\usepackage{textcomp}
\usepackage{xcolor}
\usepackage{mathtools}
\usepackage{float}
\usepackage[utf8]{inputenc}
\usepackage{xparse}
\usepackage{hyperref}
\ifCLASSOPTIONcompsoc \usepackage[caption=false,font=normalsize,labelfont=sf,textfont=sf]{subfig}
\else
\usepackage[caption=false,font=footnotesize]{subfig}
\fi

\usepackage[ruled,lined, noend,linesnumbered]{algorithm2e}
\usepackage{algorithmicx}
\SetKwIF{If}{ElseIf}{Else}{if}{}{else if}{else}{end if}%
\SetKwFor{While}{while}{}{end while}%
\SetKwFor{For}{for}{}{end for}%
\SetKwRepeat{Do}{do}{while}

\usepackage{mathtools}
\DeclarePairedDelimiter\ceil{\lceil}{\rceil}

\graphicspath{{./}}

\def\BibTeX{{\rm B\kern-.05em{\sc i\kern-.025em b}\kern-.08em
    T\kern-.1667em\lower.7ex\hbox{E}\kern-.125emX}}
    
\begin{document}
	
	\title{Accelerating Distributed ML Training via Selective Synchronization}

	\author{\IEEEauthorblockN{Sahil Tyagi}
	\IEEEauthorblockA{\textit{Indiana University Bloomington, USA} \\
	styagi@iu.edu}
	\and
	\IEEEauthorblockN{Martin Swany}
	\IEEEauthorblockA{\textit{Indiana University Bloomington, USA} \\
	swany@indiana.edu}
	}	
	
	\maketitle
	
	\begin{abstract}
		\input{abstract}
	\end{abstract}
	
	\vspace{0.1cm}
	
	
	\input{intro}
	\input{bg}
	\input{selsync}
	\input{eval}
	\input{conclusion}

\end{document}

%% file: abstract.tex
In distributed training, deep neural networks (DNNs) are launched over multiple workers concurrently and aggregate their local updates on each step in bulk-synchronous parallel (BSP) training.
However, BSP does not linearly scale-out due to high communication cost of aggregation.
To mitigate this overhead, alternatives like Federated Averaging (FedAvg) and Stale-Synchronous Parallel (SSP) either reduce synchronization frequency or eliminate it altogether, usually at the cost of lower final accuracy.
In this paper, we present \texttt{SelSync}, a practical, low-overhead method for DNN training that dynamically chooses to incur or avoid communication at each step either by calling the aggregation op or applying local updates based on their significance.
We propose various optimizations as part of \texttt{SelSync} to improve convergence in the context of \textit{semi-synchronous} training.
Our system converges to the same or better accuracy than BSP while reducing training time by up to 14$\times$.

%% file: intro.tex
\section{Introduction}\label{sec:intro}

DNNs minimize an objective function by iteratively learning the underlying relationships from some training data.
Stochastic gradient descent (SGD) \cite{b0} is the update rule that enables iterative learning based on the current loss and model parameters.
DNNs typically converge after many steps or iterations, and model quality is heavily influenced by certain variables or \emph{hyperparameters} like batch-size, number of epochs, learning rate, type of optimization (e.g., Nesterov \cite{b1}, Adam \cite{b2}, etc.), activation function, weight decay, regularization, etc.
Distributed Data-Parallel (DDP) training expedites convergence by launching model replicas over many workers concurrently and periodically synchronize their updates.
However, distributed training can suffer from high communication costs due to large model size, large number of workers and available bandwidth.
Many techniques attempt to attenuate this slowdown either by reducing the frequency or volume of communication, or relax the constraints on extent of synchronization itself, but speedup is usually achieved at the detriment of final accuracy.
We refer to such techniques that implement a loose form of synchronization or apply infrequent communication based on some heuristic as \emph{semi-synchronous}.

In this paper, we propose \texttt{SelSync} \footnote{Code available at \href{https://github.com/sahiltyagi4/SelSync}{https://github.com/sahiltyagi4/SelSync}}, a semi-synchronous training mechanism that switches between synchronous and local updates by assessing which updates are crucial enough to be communicated and which ones can be applied just locally.
It thus achieves accuracy targets akin to BSP, while reducing training time by mitigating expensive communication and applying local updates when possible.
We make the following contributions in this work:
\begin{itemize}
	\item We develop \texttt{SelSync} over the Parameter server (PS) architecture \cite{b3} using PyTorch RPC framework \cite{b37}.
	\item Implement a low-overhead technique to track changes in gradients on each step and define a threshold metric that correlates gradient changes with the trajectory of model convergence.
	Based on this threshold, we develop a communication rule that updates a model either locally or synchronously among the participating workers.
	\item Design a custom data-partitioning scheme optimal for semi-synchronous training mechanisms.
	\item Show empirically that parameter aggregation performs better than gradient aggregation in semi-synchronous training.
	This is because local and global model states tend to diverge more in the latter when workers train locally for longer and updates are aggregated infrequently.
	\item Apply \texttt{SelSync} in non-IID data settings by extending a randomized data-injection approach \cite{b35} that significantly improves model accuracy over other federated algorithms.
	\item We compare \texttt{SelSync} with other state-of-the-art algorithms across a variety of DNNs to validate the efficacy of our approach.
\end{itemize}

%% file: bg.tex
\section{Background and Related Work}\label{sec:bg}

In this section, we introduce some popular and state-of-the-art distributed learning algorithms, and briefly discuss their advantages and limitations.
Based on their respective communication patterns, each affects the parallel scaling of distributed training in a distinct manner; we refer to this as the \textit{parallel efficiency} of DDP training.
However, unlike traditional distributed applications, the computational work performed by workers at each training step is not equally important; some updates are more critical than others.
This is because the work performed by SGD does not fully compose due to the stochastic nature of gradient descent.
We refer to this as the \textit{statistical efficiency} of distributed training, and describe it in detail in \S \ref{subsec:detectcritical}.

\subsection{Bulk-Synchronous Parallel (BSP) training}\label{subsec:bsp}

In bulk-synchronous parallel or BSP training, each worker trains a DNN on a unique partition of data sampled in an IID (independent and identically distributed) manner, and aggregate their updates at the end of every training step/iteration.
Each step thus involves feeding data to every local model to predict an output, evaluate loss, compute gradients, a communication op to aggregate updates and apply them before proceeding to the next step.
DNN training is an iterative-convergent process that makes multiple passes over training dataset (where each pass is an \textit{epoch}) until loss reaches a low-enough target.
A model with parameters $w$ training over $N$ workers in parallel processes mini-batch $\mathit{x_{(i, n)}}$ of size $|\mathit{b}|$ from dataset $\mathcal{B}_{n}$ to minimize loss function $\mathcal{F}$ as:

\begin{equation}
	w_{i+1} = w_{i} - \eta \dfrac{1}{N} \sum_{n=1}^{n=N}{\dfrac{1}{|b|} \sum_{\mathit{x_{(i, n)}} \in \mathcal{B}_{n}} \dfrac{\partial}{\partial w_{i}} \mathcal{F}(x_{(i, n)},w_{i})}
	\label{eqn:distsgd}
\end{equation}

Parameters at step $(\mathit{i} + 1)$ are updated from parameters at $\mathit{i}$ in a direction opposite to the gradients, and scaled by learning rate (lr) $\eta$.
In classic BSP, updates from $N$ workers are reduced \emph{at every step}.
Communication can be performed via central parameter server (PS) \cite{b3} or Allreduce-based collectives \cite{b38}.
The aggregation step in BSP is \textit{blocking}, i.e., all workers wait for the reduction phase to complete before the next step begins.

Thus, BSP suffers from \textit{systems heterogeneity} where overall speedup is limited by the slowest worker or \emph{straggler} \cite{b0014}.
Systems heterogeneity in a cluster arises from workers with different computational capabilities.
Assuming negligible IO overhead, total iteration time mainly comprises of the computation and synchronization time: $t_{it} = t_{c} + t_{s}$.
The former is the time taken to assess loss and compute gradients, and the latter is time taken to communicate updates among workers.
ML accelerators like GPUs and TPUs accelerate computing, making computation overhead negligible (i.e., small $t_{c}$) \cite{b4}.
Thus, communication or synchronization time ($t_{s}$) tends to be the major bottleneck in scaling-out BSP training.

\begin{figure}
	\subfloat[Communication overhead]{\includegraphics[width=0.23\textwidth]{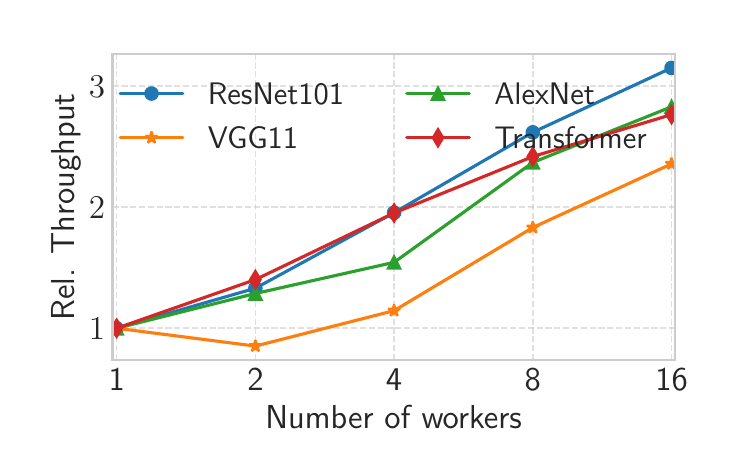}
	\label{reltput}}
	\hfill
	\subfloat[FedAvg: IID vs. non-IID data]{\includegraphics[width=0.23\textwidth]{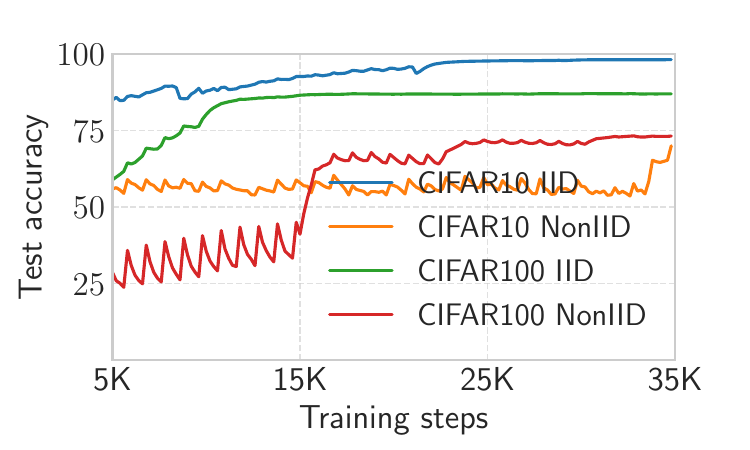}
	\label{fedavgnoniid}}
	\caption{(a) Relative throughput from training on NVIDIA V100 GPUs connected over 5Gbps NIC. (b) Training on 10 V100s with 1 label per-worker for non-IID CIFAR10 on ResNet101 and 10 labels per-worker for non-IID CIFAR100 on VGG11.}\label{fig:limits1}
\end{figure}

Due to Amdahl's law \cite{b5}, scaling thus does not increase linearly with workers due to high communication cost in DDP.
It is possible to overlap computation with communication, but the former tends to be much smaller than the latter and so the scope of parallelizing communication is limited by size of updates to exchange, available bandwidth, inter-node latency, etc. 
Fig. (\ref{reltput}) shows training throughput (i.e., number of training samples processed per second) relative to a single GPU worker as cluster-size increases with parameter server training.
Clusters with size 2 and 4 train with 1 NVIDIA V100 GPU per-node, while clusters of size 8 and 16 train with 2 and 4 GPUs per-node respectively.
Inter-node workers communicate over a 5 Gbps interface.
For ResNet101 \cite{b10}, throughput improves by only 3$\times$ even as we scale from 1 to 16 workers.
Relative throughput for VGG11 \cite{b11} at 2 workers was less than 1.0 due to high communication overhead for the 507 MB model, although it improved later over larger clusters.
We thus see that DDP throughput does not increase linearly; throughput does not improve by a factor of $N$ when training is scaled from 1 to $N$ workers.

\subsection{Federated Averaging (FedAvg)}\label{subsec:fedavg}

In federated learning (FL) \cite{b6}, models are trained over a large number of clients while keeping the data local to each.
The data on the clients may have either an IID or non-IID distribution as well.
FedAvg \cite{b7} is popular FL algorithm for scenarios where workers have limited bandwidth or data privacy concerns.
Here, updates from only a fraction of clients are aggregated \textit{infrequently} at an interval specified by the user.
Although FedAvg reduces the communication frequency considerably, it may suffer from high generalization error (i.e., lower test accuracy) due to infrequent synchronization.
Thus, there is a complex trade-off space between the communication savings and model quality (i.e., parallel and statistical efficiency) in FedAvg.
With its low-frequency, high-volume communication strategy, FedAvg collects updates from fraction $C$ of $N$ workers $x$ times in every epoch, such that synchronization factor $E$=$1/x$.
For e.g., $C$=0.5 with $E$=0.25 implies parameters are synchronized among half the devices 4 times (uniformly spaced) over each epoch.
One of the key objectives of FL is to improve convergence on workers with non-IID data.
This is especially common in mobile/edge computing where data is stored in a decentralized manner across workers.
The performance difference from training on balanced vs. unbalanced data with FedAvg is shown in Fig. (\ref{fedavgnoniid}) from running ResNet101 and VGG11.
We train on a cluster of 10 V100 GPUs such that non-IID versions of CIFAR10 (with a total of 10 labels) and CIFAR100 (with 100 labels) \cite{b8} are partitioned as 1 and 10 labels per-worker respectively.
We set $C$=1 and $E$=0.1 so updates were collected from all workers 10 times in an epoch (every 1/10\textit{th} epoch).

\subsection{Stale-Synchronous Parallel (SSP)}\label{subsec:ssp}

\begin{figure}
\subfloat[Compute time vs. batch-size]{\includegraphics[width=0.23\textwidth]{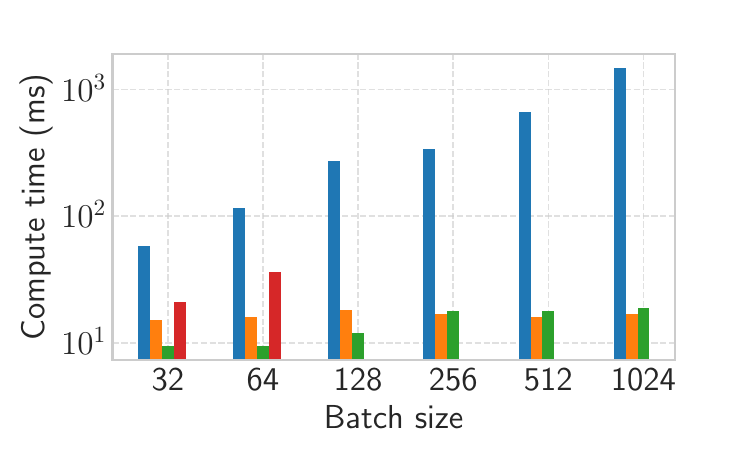}
\label{sspcomputetime}}
\hfill
\subfloat[Memory util. vs. batch-size]{\includegraphics[width=0.23\textwidth]{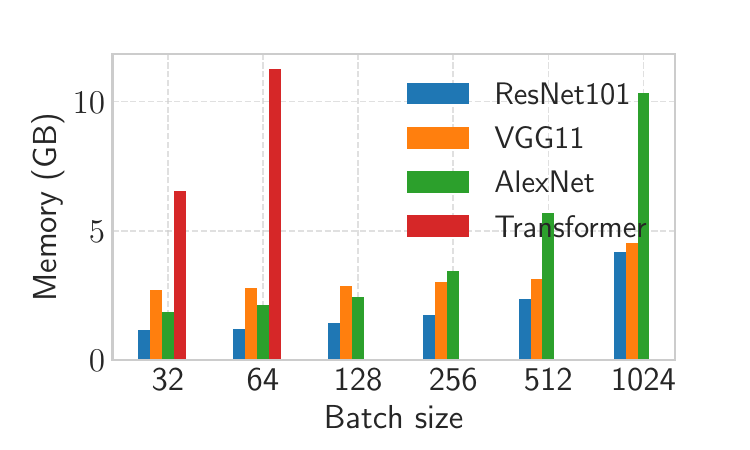}
\label{sspmemory}}
\caption{Setting worker batch-size to $Nb$ in SSP ensures same amount of work is done at each iteration as in BSP. However, computation and memory requirements also increase with batch-size, as measured on a Tesla K80 GPU.}\label{fig:ssplimits}
\end{figure}

SSP \cite{b47} is a PS-based approach to mitigate systems heterogeneity between workers which train asynchronously and independently update the global parameters on the central PS in a non-blocking manner.
However, the asynchronicity is conditional; synchronization is enforced by blocking faster workers when they exceed the slower workers by a user-specified threshold.
Once the threshold exceeds, faster workers are forced to halt and wait for slower workers to catch up.
For e.g., setting \textit{staleness threshold} $s$=100 implies fast workers cannot exceed slow workers by more than 100 iterations.
Thus, SSP combines non-blocking execution of asynchronous training while still ensuring the divergence of model parameters between slow and fast workers stays within bounds, determined by $s$.
Compared to fully asynchronous training, this reduces staleness effects and inconsistent model states among workers to an extent.
In BSP, each of the $N$ workers processes mini-batch size $b$ in a single step.
As updates are aggregated at each step, total computational work done in BSP corresponds to its global batch-size $Nb$.
Comparatively, SSP performs lesser work per-step as updates computed from a worker over batch-size $b$ are sent asynchronously to the PS.
One could theoretically increase worker batch-size to $Nb$ so as to perform the same amount of work as BSP.
However, this increases both the computational and resource requirements on a worker, shown in Fig. (\ref{fig:ssplimits}).
The computation time across different batch sizes on an NVIDIA K80 is shown in Fig. (\ref{sspcomputetime}) for various image and language models
Correspondingly, Fig. (\ref{sspmemory}) shows how memory consumption on a worker shoots up with the batch-size.
Transformer \cite{b9} on WikiText-103 \cite{b14} failed to scale beyond $b$ = 64 due to OOM (out-of-memory) error as memory requirements exceeded the GPU's 12GB capacity.
AlexNet \cite{b12} training on the large ImageNet-1K \cite{b13} dataset had high memory utilization with large $b$ when using PyTorch's \texttt{torchvision.datasets.ImageFolder} for data loading and batching \cite{b009}.
Although relatively smaller in size, ResNet101 is the deepest of all the models evaluated here (with 101 layers).
Thus, feeding more samples with larger $b$ increases the forward passes made through the layers of the network which in turn increases computation time on larger batches.
If cluster-size $N$ is too large, setting per-worker batch-size to $Nb$ would greatly increase the compute time and memory on each worker, thus making SSP impractical or even prone to failures in those scenarios.

\textbf{Summary:} \textit{DDP methods either maximize \textit{useful} work by iteratively aggregating worker updates (BSP), or speedup training by reducing communication frequency (FedAvg), or loosen the constraints on synchronization itself (SSP).
The latter two attain considerable speedup over BSP when they work, but see significant drop in final accuracy/loss when they don't.
This is because they only consider the systems aspect and \textbf{not} the statistical aspect of parallelizing DNN training.
We describe in \S \ref{subsec:deltaselsync} on how we attempt to bridge this gap by designing a training strategy that considers both.}

\begin{figure}
	\subfloat[ResNet101 at epoch 1]{\includegraphics[width=0.23\textwidth]{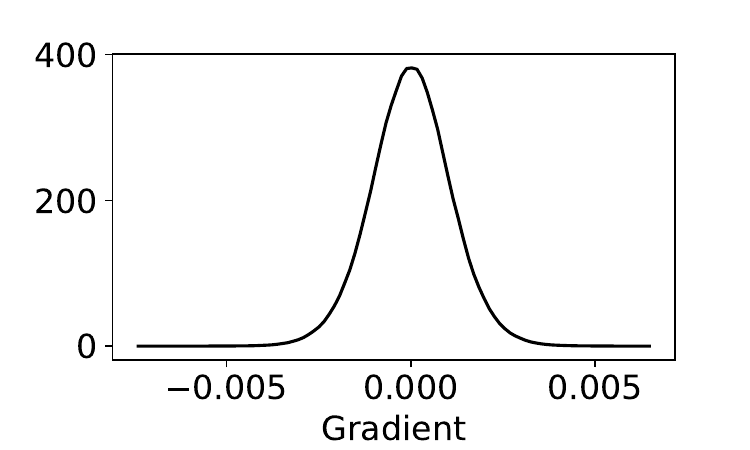}}
	\hfill
	\subfloat[ResNet101 at epoch 50]{\includegraphics[width=0.23\textwidth]{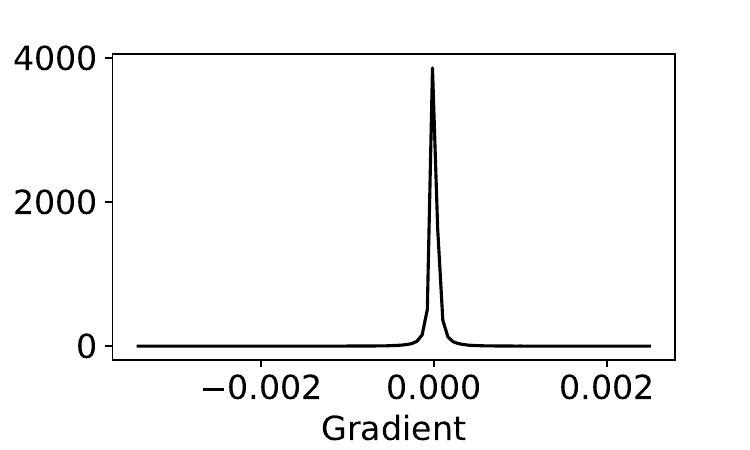}}
	\hspace{0.1cm}
	\subfloat[Transformer at epoch 1]{\includegraphics[width=0.23\textwidth]{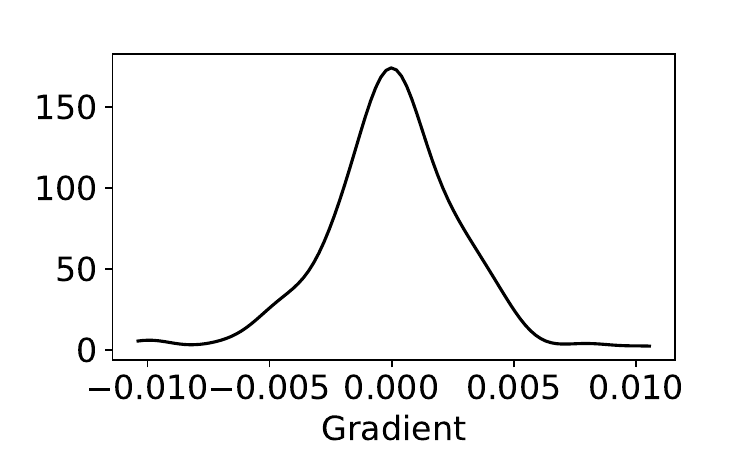}}
	\hfill
	\subfloat[Transformer at epoch 4]{\includegraphics[width=0.23\textwidth]{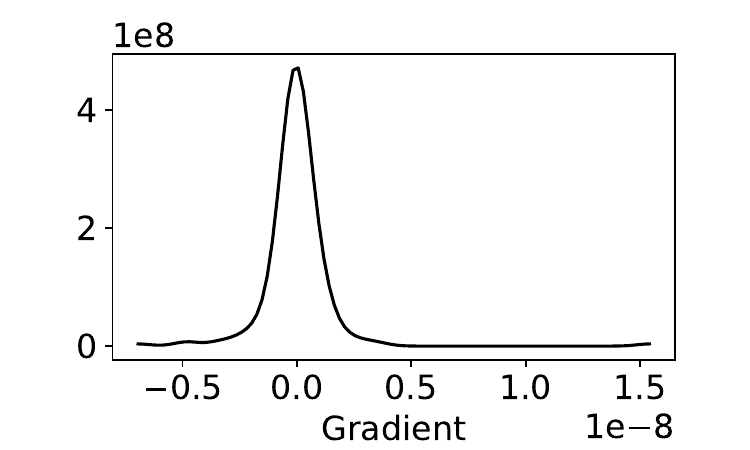}}
	\caption{Gradient Kernel density estimates (or KDE on the Y-axis) over the epochs for ResNet101 layer \texttt{layer4\_1\_conv1\_weight} and Transformer layer \texttt{transformer\_encoder\_layers\_0\_norm1\_weight}. Gradients are volatile in the early epochs ((a) and (c)), but get smaller and saturate as gradient variance decreases over the course of training ((b) and (d)).}
	\label{fig:bggraddist}
\end{figure}

\subsection{Other communication-optimizing DDP methods}\label{subsec:commoptmethods}

Gradient compression techniques lower the volume or density of gradient tensors based on some rule or heuristic.
They can broadly be classified into: \emph{sparsification}, \emph{quantization} or \emph{low-rank approximations}.
We send a subset of gradient values and its corresponding indices in sparsification, effectively reducing the volume of gradients to exchange, as done in DGC \cite{b28} and Top-\textit{k} \cite{b26}.
In quantization, we map the gradients to a finite set of values or reduce the bit-width of single-precision floating point gradients, e.g., signSGD \cite{b50}, Terngrad \cite{b49}, etc.
Low-rank techniques like PowerSGD \cite{b51} decompose gradient tensors into low-rank matrices, thus reducing the density of gradient tensors to synchronize.

Compression is not a zero-cost operation; there can be considerable overhead in compressing and decompressing gradients \cite{b0026}.
\emph{Although a high compression factor may improve overall throughput, it may degrade the final model quality (compared to BSP) or require more training (thus, increasing end-to-end training time).}
On the other hand, pipeline-parallel frameworks like PipeDream \cite{b41} do not eliminate communication, but interleave computation with communication by breaking training mini-batches into micro-batches.
This approach keeps the execution pipeline busy and avoids idling by calculating gradients on the next micro-batch while that of the current micro-batch are exchanged.
GradientFlow \cite{b42} also overlaps computation with communication by exploiting the sequential execution nature of backpropagation, communicating outer layers of a neural network while concurrently computing gradients of the inner layers over backward pass.
In addition, GradientFlow achieves 2$\times$ compression with FP16 mixed-precision training \cite{b43} and aggregates via bandwidth-optimal ring-allreduce algorithm.
Similarly, ByteScheduler \cite{b44} optimizes communication schedule further by repartitioning and rearranging tensors for efficient transmission.
Instead of synchronizing each of the smaller layers separately, ByteScheduler thus batches their communication as a single operation.
\emph{Rather than eliminating communication overhead altogether, these methods either optimize the communication schedule or reduce the communication volume to synchronize among workers.}

\subsection{Gradient sensitivity through DNN training}\label{subsec:gradchange}

\begin{figure}
	\subfloat[ResNet101]{\includegraphics[width=0.23\textwidth]{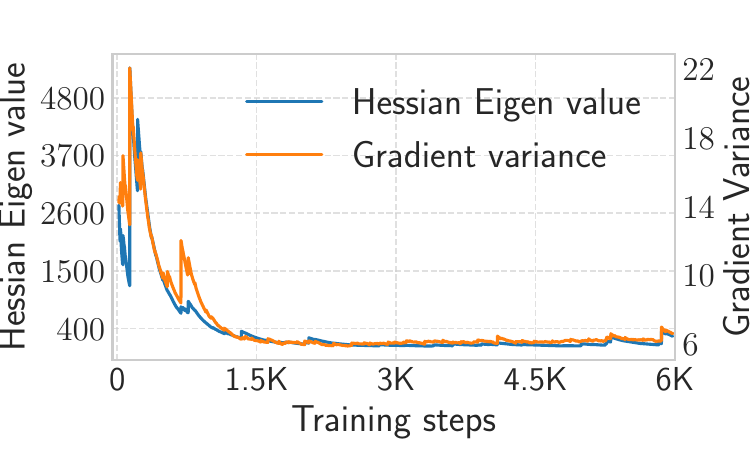}
	\label{rensethessian}}
	\hfill
	\subfloat[VGG11]{\includegraphics[width=0.23\textwidth]{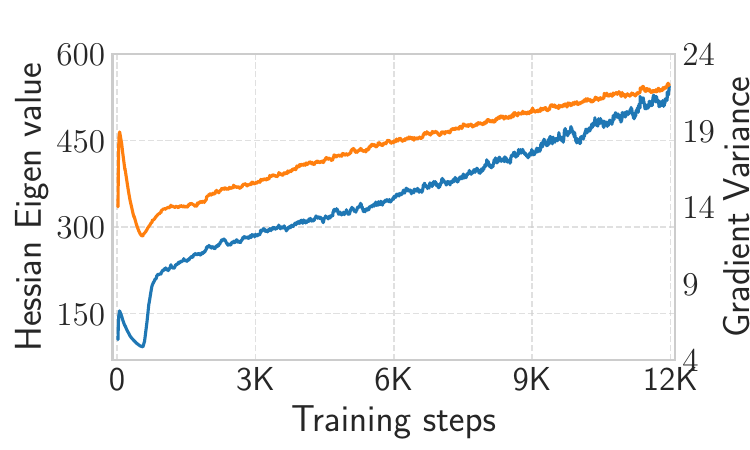}
	\label{vgghessian}}
	\caption{We compute the largest eigenvalue of second-order Hessian on every iteration along with first-order gradient variance.
	Changes in eigenvalue of the Hessian help detect critical periods, and it can be approximated with the variance of first-order gradients.
	The latter has a significantly lower computational overhead.}\label{fig:hessianeigenval}
\end{figure}

In early stages of DNN training, gradients computed during backpropagation are large and change drastically \cite{b15}.
This is because a randomly initialized model adjusts its parameters aggressively towards the minima in early training phases.
With optimal hyperparameters, a model would eventually converge after many iterations.
Thus, updates computed in later stages become smaller as a DNN approaches convergence and gradients stop changing over time.
We sample a mini-batch $b$ and update the DNN from the gradients computed from it via Mini-batch Gradient Descent.
However, updates computed from a randomly sampled mini-batch are not representative of the true gradients computed over the entire training data and tend to be noisier \cite{b19}.
Fig. (\ref{fig:bggraddist}) shows how gradients vary throughout training by plotting kernel density estimates (KDE) of the gradients for a layer of ResNet101 and Transformer over different epochs.
The training routine for the models is described in \S \ref{subsec:clustersetup}.
As training progresses, gradients become smaller and more gradients are closer to 0 at epoch 50 and 4 vs. epoch 1.
Recent studies have shown that gradients are not only sensitive in the early phases, but also at certain \emph{critical} or \emph{sensitive regions} \cite{b17,b18}, and heavily influenced by training hyperparameters (like learning-rate schedule, gradient clipping and SGD variant).
Hessian information (i.e., second-order gradients) helps measure the divergence in gradients for a given training batch-size \cite{b48}.
\cite{b18} observed that changes in the eigenvalues of the Hessian works as an effective indicator of critical learning periods, while \cite{b24} verified that first-order gradient norm approximates to the inter-iteration Hessian eigenvalues for ResNet18 on CIFAR10.
We validate the same for larger models like ResNet101 and VGG11, shown in Fig. (\ref{fig:hessianeigenval}) and see how the largest eigenvalue of the Hessian computed over each training step follows a similar course as first-order gradient variance.
However, the latter is much more cheaper to compute and can be integrated into the backpropagation routine. 
Even though their magnitudes lie on different scales, the relative inter-iteration changes are similar.
By this approximation, \textit{gradient noise} or \textit{variance} helps account for the statistical efficiency of DNN training that varies with batch-size, type, size/complexity of a DNN and its hyperparameters.

\textbf{Summary:} \textit{Gradients get smaller and eventually saturate, but the exact magnitude and trajectory of gradient progression varies for a DNN, dataset and DNN-specific hyperparameters.}

%% file: selsync.tex
\section{Design and Implementation}\label{sec:selsync}

In this work, we leverage both the parallel and statistical aspect of DDP training by developing a flexible communication strategy to perform either synchronous or local-SGD updates based on their significance.
Parallel efficiency is improved via local-SGD where workers apply their respective updates locally and avoid communication which reduces iteration time and speeds up training.
When gradients change significantly, statistical efficiency is prioritized by synchronizing updates among all workers.
Although this increases the iteration time due to communication, the aggregated gradients are less noisy than the locally computed gradients.
These averaged updates are then applied globally to produce a model state consistent across all workers.

\subsection{Detecting crucial updates in DNN training}\label{subsec:detectcritical}

Distributed training extends metrics like \emph{throughput} and \emph{scale-out factor} from fields like HPC and parallel computing to measure the parallel efficiency of collaboratively training a model across various workers.
However, DNN training has a statistical aspect as well that is \emph{non-linear}; some updates are more critical than others towards overall model learning and generalization.
As seen previously in \S \ref{subsec:gradchange}, gradients change considerably in the early phase and at certain critical regions.

Recent works have demonstrated that gradient variance, or signal-to-noise ratio between the local and aggregated updates can serve as a useful indicator of the statistical efficiency of DDP training \cite{b20,b21,b22}.
In fact, there is a strong correlation between the changes in the eigen values of the second-order Hessian \cite{b23} and the norm of first-order gradients \cite{b24}.
Computing gradient norm is computationally feasible as it can easily be incorporated in backpropagation, while computing the Hessian is very expensive from a computational standpoint.
We measure the significance of each update by tracking changes in gradient norm on every iteration.
\textbf{\textit{Relative gradient change}} between consecutive steps can be estimated from the $\mathbb{L}{2}$-norm of gradient $\nabla\mathcal{F}$ at steps $(\mathit{i}-1)$ and $\mathit{i}$, as shown in Eqn. (\ref{eqn:gradchange}):

\begin{equation}
	\triangle(g_{i}) = \bigg|\dfrac{\mathbb{E}[||\nabla\mathcal{F}_{(i)}||^{2}] - \mathbb{E}[||\nabla\mathcal{F}_{(i-1)}||^{2}]}{\mathbb{E}[||\nabla\mathcal{F}_{(i-1)}||^{2}]}\bigg|
	\label{eqn:gradchange}
\end{equation}

\begin{figure}
	\subfloat[ResNet101]{\includegraphics[width=0.23\textwidth]{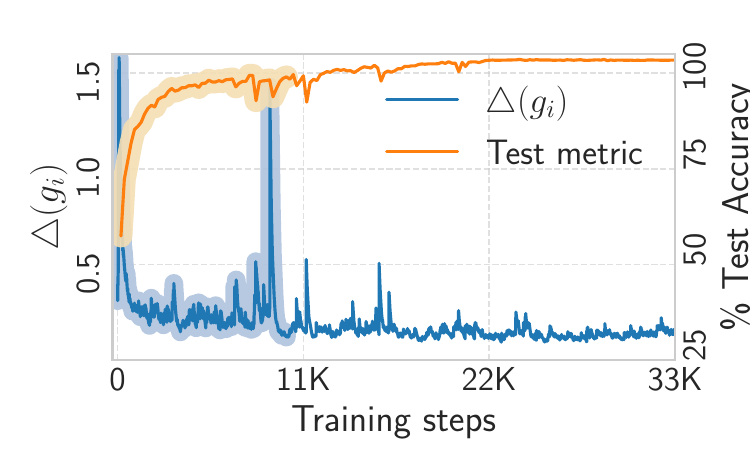}
	\label{resnetgradchangeacc}}
	\hfill
	\subfloat[VGG11]{\includegraphics[width=0.23\textwidth]{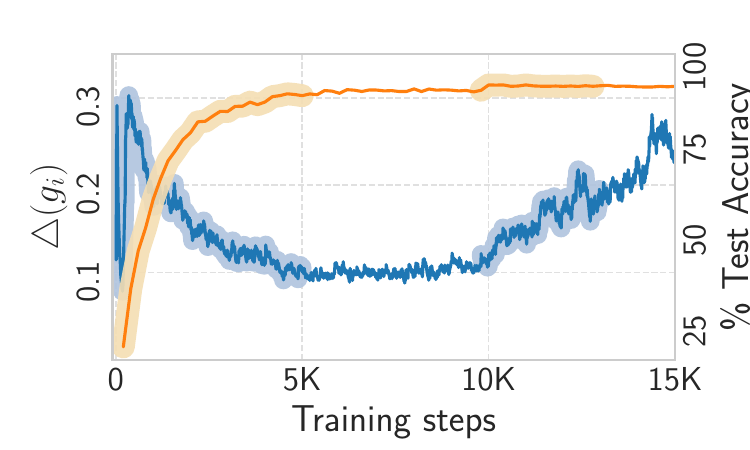}
	\label{vgggradchangeacc}}
	\hspace{0.1cm}
	\subfloat[AlexNet]{\includegraphics[width=0.23\textwidth]{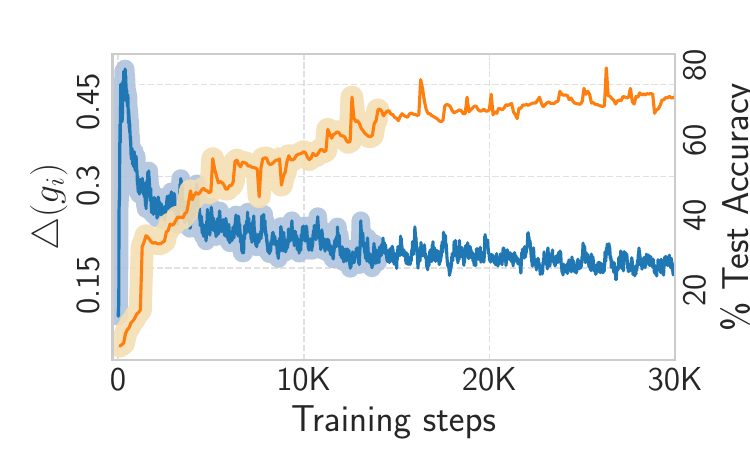}
	\label{alexnetgradchangeacc}}
	\hfill
	\subfloat[Transformer]{\includegraphics[width=0.23\textwidth]{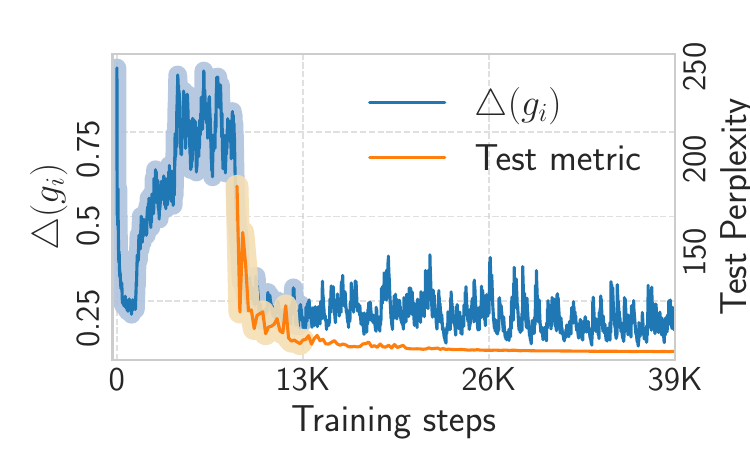}
	\label{transformergradchangeacc}}
	\caption{Correlation between relative gradient change and model convergence in BSP.
	A rise or decline in test accuracy or perplexity is accompanied by changes in the gradients as well.
	As convergence plateaus, so does $\triangle (g_{i})$.}
	\label{fig:gradchangeaccuracy}
\end{figure}

Gradients computed on a single iteration are noisy, so we use exponentially weighted moving average (EWMA) \cite{b39} for smoothing.
We plot $\triangle(g_{i})$ at every iteration $i$ alongside the convergence curves in Fig. (\ref{fig:gradchangeaccuracy}).
The curves correspond to test accuracy for ResNet101, VGG11 and AlexNet (the higher, the better), and test perplexity for Transformer (the lower, the better).
We apply EWMA with a window-size of 25 iterations and a smoothing factor of $N$/100, or 0.16 for a 16-node cluster.
The plots highlight the correlation between critical updates (measured by $\triangle(g_{i})$) and model progression during training.
The particularly volatile phases are shown as shaded in Fig. (\ref{fig:gradchangeaccuracy}).
For Transformer, we omit perplexity for first 8K iterations as it was very high initially and fell sharply in that phase.

Fig. (\ref{resnetgradchangeacc}) and Fig. (\ref{transformergradchangeacc}) shows how gradients change in ResNet101 and Transformer as the accuracy/perplexity changes sharply.
The sudden peak in ResNet101 after 10K steps corresponds to learning rate decay.
The first 10K steps in Transformer where the perplexity fell aggressively, the associated gradients fall and rise sharply as well.
The perplexity falls gradually after 13K iterations, as does the corresponding gradient change.
As for AlexNet, Fig. (\ref{alexnetgradchangeacc}) shows how $\triangle(g_{i})$ varies gradually such that accuracy gains over the same set of iterations are gradual as well.
The gradients for VGG11 in Fig. (\ref{vgggradchangeacc}) change rapidly in the first 5K steps as accuracy improves steeply in that phase.
As accuracy saturates between iterations 5K and 10K, gradients stop changing as well and $\triangle(g_{i})$ flattens out in that phase.
The accuracy then jumps up after 10K\textit{th} step due to learning rate decay alongside $\triangle(g_{i})$.

\subsection{$\delta$-based selective synchronization}\label{subsec:deltaselsync}

Gradient changes tend to be more prominent with aggressive updates, and stabilize over the course of training.
As a model converges, gradients stop changing as abruptly, as does $\triangle(g_{i})$.
The maximum degree by which gradients vary depends on DNN size and complexity, weight initialization, among other hyperparameters.
Denoting this extremum as $\mathcal{M}$=$max([\triangle(g_{i})]_{i=1}^{i=I})$ after running for $I$ iterations, gradient change then lies in the range $[0, \mathcal{M}]$.

\begin{figure}
  \includegraphics[width=0.45\textwidth]{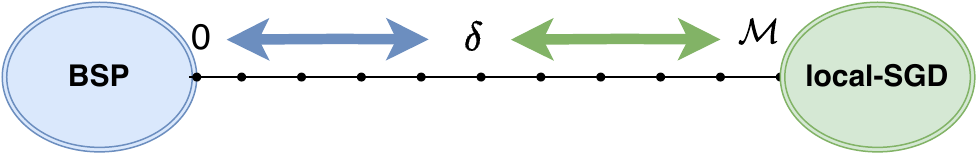}
  \caption{Adjusting threshold $\delta$ on relative gradient change. Choose BSP if $\triangle(g_{i}) \geq \delta$ and local SGD if $< \delta$. Setting $\delta$=0 implies BSP training, while a very high $\delta$ trains only with local updates.}
  \label{fig:deltaslider}
\end{figure}

With this intuition, it is thus logical to synchronize worker updates only when $\triangle(g_{i})$ changes considerably, and perform local updates to avoid communication otherwise.
By choosing perceptively between synchronous and local updates, we try to balance the parallel and statistical efficiency of DDP training.
Additionally, incorporating local-SGD updates in otherwise synchronous training allows for \emph{more} exploration of the local minima by every worker which in turn can improve model generalization \cite{b33, b40}.
By allowing workers to train locally, \texttt{SelSync} thus allows model replicas to explore their local latent space, while still bounding the divergence between local and global model states.
The latter is located on the central PS node.
\texttt{SelSync} proposes an effective and practical method to measure the significance of inter-iteration updates by imposing a threshold, denoted by $\delta$, on relative gradient change described as follows: \textit{use local-SGD updates if gradient change is below $\delta$ and synchronize updates if \textbf{any} of the workers' $\triangle(g_{i}) \geq \delta$}.
Threshold `$\delta$' is set prior training launch such that $\delta$=0 implies fully synchronous training while setting a sufficiently large threshold ($\delta \geq \mathcal{M}$) trains with local-SGD only.
In case of fully local-SGD training, workers never communicate with each other and explore only their respective local minimas.
Fig. \ref{fig:deltaslider} illustrates how one can slide around $\delta$ between 0 and $\mathcal{M}$ to adjust the degree of training between synchronous and local updates with \texttt{SelSync}.

\subsection{Gradient vs. Parameter aggregation for \texttt{SelSync}}\label{subsec:gradvsparamagg}

In BSP, synchronization can either be performed on the \textit{gradients} or \textit{model parameters}.
They are equivalent in BSP \emph{assuming all workers initially started with the same model parameters}.
On PS, this is done by workers pulling initial model state from the central server prior training, while decentralized topologies can broadcast model parameters from one of the workers to all others in the cluster.
Local models are thus consistent with the global state on every step in BSP as the same, aggregated set of updates are applied to each worker.
In spite of their equivalence, gradient aggregation is more popular as gradients can be compressed \cite{b25,b0026} via sparsification \cite{b26,b27,b28}, quantization \cite{b29,b30} or low-rank updates \cite{b31}.
Model parameters can be compressed via pruning \cite{b32} as well, although model generalization and compression factor achieved is comparatively subpar.

However, \textit{gradient aggregation (GA) and parameter aggregation (PA) are not equivalent in the context of semi-synchronous training}; reducing model parameters instead of gradients tends to produce DNNs with better generalization performance.
Since \texttt{SelSync} collects updates infrequently (depending on $\triangle(g_{i})$ and chosen $\delta$), GA can cause local models to diverge from the global state on the PS and thus, lead to model inconsistency across workers.
The more the iterations are performed locally, the more local replicas can diverge as a result.
This is because even though the same set of averaged gradients are applied to each worker in GA, the resultant parameters post-SGD update will be different across workers and may eventually become inconsistent, especially after workers have been training locally for many subsequent iterations.
This increased divergence can thus deteriorate the test performance in semi-synchronous methods.

However, local training is not always bad; lettings workers train with local-SGD permits ample exploration of the local minima \cite{b33}.
Adding PA on top of that ensures local weights are bounded to the global weights and do not fall into a local minima that is too far from the global state.
Further, aggregating parameters during synchronization phase ensures model consistency across all workers.
We empirically validate that PA achieves same or better convergence than GA across a variety of DNNs when training with  \texttt{SelSync}.

\subsection{IID data partitioning for semi-synchronous training}\label{subsec:iiddataselsync}

\begin{figure}
	\centering
	\subfloat[Default data-partitioning (DefDP) in a 4 worker cluster.]{{\includegraphics[width=0.4\textwidth]{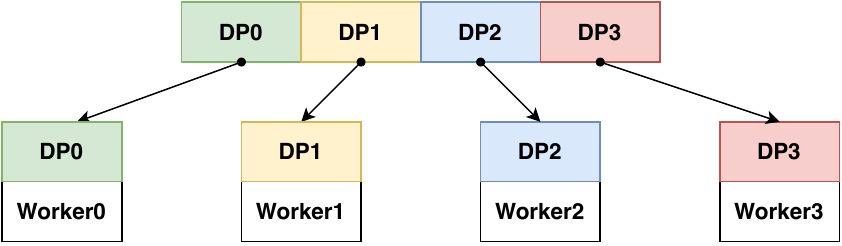}}\label{fig:defaultDP}}
	\vspace{0.25cm}
	\subfloat[SelSync data-partitioning (SelDP) in a 4 worker cluster.]{{\includegraphics[width=0.4\textwidth]{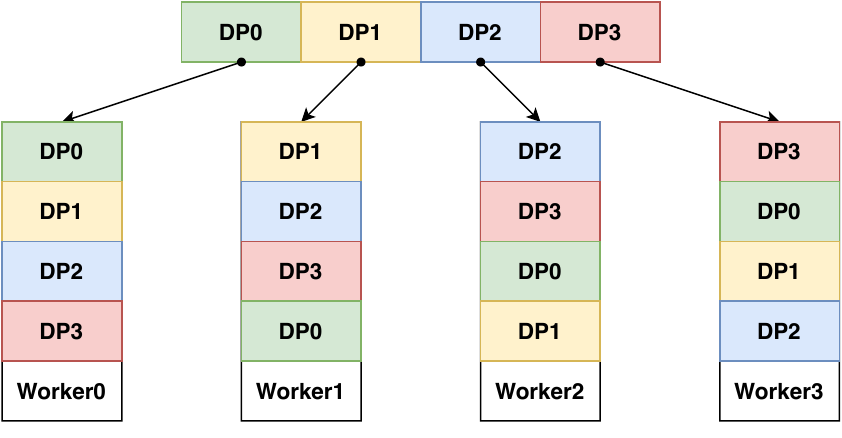}}\label{fig:selsyncDP}}
	\caption{Data-partitioning (DP) in distributed training.
	DefDP splits training data into as many partitions as the number of workers.
	SelDP partitions data as a circular queue whose head is rotated to a unique chunk on each worker.}
	\label{fig:partitionscheme}
\end{figure}

BSP maximizes per-iteration work by processing a unique mini-batch on every worker and combining updates from those partitions.
For example, training dataset $\mathcal{D}$ can be partitioned among 4 workers as $\mathcal{D} \supseteq \{$\texttt{DP0}, \texttt{DP1}, \texttt{DP2}, \texttt{DP3}$\}$, such that each partition resides on a single worker.
We refer to this as \textit{Default data-partitioning} or \textit{DefDP}, depicted in Fig. (\ref{fig:defaultDP}) where \texttt{worker0} samples from partition \texttt{DP0}, \texttt{worker1} from \texttt{DP1}, \texttt{worker2} from \texttt{DP2} and \texttt{worker3} from \texttt{DP3}.
This approach works well in BSP as every training sample is used at least once in each epoch to update model parameters.

However, such a partitioning scheme does not work well for semi-synchronous methods.
Consider a DNN training with \texttt{SelSync} where most of the iterations favor local-SGD over synchronous updates.
Every worker then optimizes the model for its local training subset only and fails to aggregate updates from other workers computed on different data partitions.
Thus, local replicas are deprived of learning features from data partitions situated on other workers due to low or negligible communication.
Even if aggregation is triggered occasionally, averaging diverge model can produce a resultant model of low quality \cite{b7}.
Training DNNs with DefDP in semi-synchronous methods is somewhat analogous to training with unbalanced data in federated learning, which is detrimental to model convergence as well.

As part of \texttt{SelSync}, we propose a custom partitioning scheme tailored for techniques using a hybrid of local and synchronous updates.
Instead of splitting data into unique and separate chunks, we shuffle dataset $\mathcal{D}$ on every worker based on its ID in the cluster (in case of decentralized training using allreduce collectives, this corresponds to a worker's rank).
So, \texttt{worker0} computes on partitions ordered as $\{$\texttt{DP0}, \texttt{DP1}, \texttt{DP2}, \texttt{DP3}$\}$, \texttt{worker1} processes in the order $\{$\texttt{DP1}, \texttt{DP2}, \texttt{DP3}, \texttt{DP0}$\}$, \texttt{worker2} as $\{$\texttt{DP2}, \texttt{DP3}, \texttt{DP0}, \texttt{DP1}$\}$, while \texttt{worker3} samples batches ordered as $\{$\texttt{DP3}, \texttt{DP0}, \texttt{DP1}, \texttt{DP2}$\}$.
The shuffling operation is a one-time overhead that is executed prior training.
Thus, \textit{SelSync data-partitioning}, or \textit{SelDP} ensures all training samples are available to every worker to train on.
This partitioning scheme is illustrated in Fig. (\ref{fig:selsyncDP}).
\textit{In this way, local model replicas will not be skewed from training with local-SGD for most iterations (say for \texttt{SelSync} with a high $\delta$-setting)}.
\textit{On the other hand, when worker updates do synchronize, SelDP ensures each workers' updates comes from a unique chunk and none of the workers redundantly process the same set of training samples for a given iteration}.
For e.g., if the very first iteration qualifies for synchronization, then \texttt{worker0} processes a mini-batch from chunk \texttt{DP0}, \texttt{worker1} processes from \texttt{DP1}, \texttt{worker2} processes from \texttt{DP2} and \texttt{worker3} processes from \texttt{DP3}.

\subsection{Training on unbalanced and non-IID data}\label{subsec:noniiddesign}

As seen in \S \ref{subsec:fedavg}, distributed training suffers from model deterioration due to unbalanced and skewed data on account of its non-IID\textit{ness}.
To mitigate this issue, we extend \textit{randomized data-injection} \cite{b35} and apply it in the context of semi-synchronous training when \textit{dealing with non-IID data}.
In data-injection, a random subset of workers communicate amongst themselves and share partial training data.
Privacy risk is minimized by sharing data with only a fraction of workers randomly and on a per-iteration basis.
Data-injection is configured as a tuple of ($\alpha, \beta$), where $\alpha$ is the fraction of workers chosen while $\beta$ is the proportion of worker batch-size to be shared.
For e.g., an ($\alpha,\beta$) of (0.5, 0.5) implies half of the total workers are selected randomly to share half of their training samples, thus improving the overall data distribution in DDP training.

The mini-batch size is an important hyperparameter in DNN training.
Naively applying data-injection can be harmful if the cluster-size is large.
This is because the per-worker batch-size would then become undesirably large as well and degrade model generalization as commonly seen in large-batch training \cite{trainlong}, \cite{b36}.
To avoid this, we enforce the following constraint in \texttt{SelSync}: when dealing with on non-IID data, adjust worker batch-size to $b^{'}$ such that cumulative batch-size after data-injection is same as the originally set batch-size $b$, shown in Eqn. (\ref{eqn:noniidinject}) for a cluster of $N$ workers:

\begin{equation}
	b^{'} = \frac{b}{(1 + \alpha \beta N)}
	\label{eqn:noniidinject}
\end{equation}

The communication overhead of data-injection is low; each iteration additionally communicates only ($\alpha \beta N b'$) times the size of an individual sample.
For CIFAR10/100, each image is about 3Kb (kilobytes), and 10-150Kb in ImageNet-1K.
Thus, data-injection with 16 workers and $b$ = 32 in a (0.5, 0.5) configuration incurs a transfer of 132Kb on CIFAR10/100 and 440-6600Kb with ImageNet-1K.
This cost is negligible compared to the cost of exchanging model updates that span hundreds of megabytes or even gigabytes in modern DNNs.
To apply data-injection in \texttt{SelSync}, we additionally append the $\delta$-threshold and denote a configuration as ($\alpha,\beta,\delta$).
Privacy risk with data-injection is reduced due to $K$-anonymity \cite{b45} (for a cluster-size of `$K$' workers).
Since workers are chosen randomly at each iteration, privacy is preserved by not knowing from which of the $\ceil*{\alpha\cdot K}$ from the total $K$ workers did the training samples came from for the current iteration.

\textbf{Training with \texttt{SelSync}}:
With the target to achieve same or better convergence as BSP while avoiding expensive communication and thus accelerate training, we develop \texttt{SelSync} for DNN training over both IID and non-IID data distributions, described in detail under Alg. (\ref{alg:selsyncalgo}).
We fashion \texttt{SelSync} over parameter server (PS) architecture where worker updates are coordinated through a central server.
But \texttt{SelSync} will also work in a decentralized topology where workers communicate amongst each other using MPI-based collectives.
Training begins at line 3 of Alg. (\ref{alg:selsyncalgo}) with the workers pulling initial model state from the PS using \texttt{pullFromPS} function.

\begin{algorithm}
	\label{alg:selsyncalgo}
	\DontPrintSemicolon
	\SetKwInOut{Input}{Input}
	\SetKwProg{Pn}{procedure}{:}{\KwRet}
	\SetKwFunction{train}{train}
	\SetKwFunction{flag}{flags}
	\SetKwFunction{bit}{bit}
	\SetKwFunction{pull}{pullFromPS}
	\SetKwFunction{push}{pushToPS}
	\SetKwFunction{trackgrad}{RelativeGradChange}
	\SetKwFunction{syncstatus}{allgather\_status}
	\caption{\{\textbf{Sel}\}ective \{\textbf{Sync}\}hronization}
	\textbf{Input:} learning rate $\eta$, gradient change threshold $\delta$, cluster-size $N$, training data $\mathcal{D}_{n}$ on worker with id $n$\\
	\Pn{\train{}}{
		$w_{(n,0)}$ = \pull{} \Comment{\textcolor{gray}{initialize parameters}}\\
		\For {$i$=0,1,...$I$ \text{on worker id $n$} \Comment{\textcolor{gray}{training iterations}}} {
			\bit[$N$] \flag = 0 \Comment{\textcolor{gray}{synchronization status}}\\
			$d_{(i,n)} \in \mathcal{D}_{n}$ \Comment{\textcolor{gray}{sample mini-batch from data}}\\
			$g_{i} = \nabla\mathcal{F}(x_{(i,k)}, w_{(n,i)})$ \Comment{\textcolor{gray}{compute gradient at $i$}}\\
			\BlankLine
			$\triangle(g_{i})$ = \trackgrad($||g_{i}||^{2}$)\\
			$w_{(n,i+1)} = w_{(n,i)} - \eta \cdot g_{i}$ \Comment{\textcolor{gray}{apply local updates}}\\
			
			\If{$\triangle(g_{i}) \geq \delta :$}{
				\flag[$n$] = 1 \Comment{\textcolor{gray}{synchronize called by worker $n$ as its gradient change exceeds $\delta$}}\\
			}
			\BlankLine
			\flag = \syncstatus{\flag} \Comment{\textcolor{gray}{call all-gather on \flag such that index $n$ holds worker $n$'s synchronization status bit}}\\
			\BlankLine
			\If{$1 \in \flag :$}{
				\push{$w_{(n,i+1)}$} \Comment{\textcolor{gray}{push local updates}}\\
				$w_{(n,i+1)}$ = \pull{} \Comment{\textcolor{gray}{pull global}}
			}
		}
	}
\end{algorithm}

Given a mini-batch sampled from dataset $\mathcal{D}_{n}$ on worker $n$, we compute the local gradients and its variance, as shown in line 7 and 8.
The relative gradient change is measured by keeping track of gradient variance via \texttt{RelativeGradChange} in the algorithm.
This function calculates and stores gradient variance at every iteration, applies EWMA smoothing, and computes the relative gradient change as shown in Eqn. (\ref{eqn:gradchange}).
The local gradients are then used to update model parameters locally (line 9).
To keep track of $\triangle (g_{i})$ \textit{across all workers}, we initialize a \texttt{flags} array of length $N$ (for the total workers in the cluster).
All elements of the array are reset to 0 at the beginning of each iteration (line 5); a bit value of 0 at index $n$ in \texttt{flags} implies \emph{$\triangle (g_{i})$ of worker with id $n$ is below threshold $\delta$ and thus, wants to perform SGD update only locally}.
Similarly, a bit value of 1 at index $n$ in \texttt{flags} means worker $n$'s \emph{relative gradient change exceeds the threshold and wants to communicate its update with other workers}.
We refer to this as the synchronization status of a worker.
Lines 10-11 show that if $\triangle (g_{i})$ on worker $n$ is greater than $\delta$, we flip the corresponding bit at index $n$ to 1.
We then call \texttt{all-gather} to communicate each worker's synchronization bit to every other worker in the cluster (line 12).
If any synchronization bit in the \texttt{flags} array is 1 post \texttt{all-gather}, we communicate gradients among the workers.
\textit{Since we only exchange a single bit corresponding to each worker, the total communication volume of this step is just $(N-1)$ bits on every worker (in a cluster of size $N$).
This op had a negligible overhead in our evaluation ($\approx$ 2-4 ms).}
If any of the bits in \texttt{flags} is 1, we synchronize by pushing local parameters of \textit{every} worker to the PS via \texttt{pushToPS} function, and pulling updated model parameters: $w_{(n,i+1)} = \sum_{n=1}^{N} w_{(n,i)} / N$, i.e., perform \textit{parameter aggregation}.
Thus, local replicas remain consistent with the global model state as they all synchronize if even a single worker chooses synchronization over local-SGD.
On the other hand, we train with local-SGD if all the bits of \texttt{flags} array are zero for a given iteration.

The dataset $\mathcal{D}_{n}$ at line 6 in Alg. (\ref{alg:selsyncalgo}) is treated differently based on its IID or non-IID properties.
When training on IID data, we implement \textit{SelDP partitioning} for \texttt{SelSync} by reordering and partitioning data on worker $n$ by first splitting training data $\mathcal{D}$ into $N$ equal-sized chunks.
SelDP uses these partitions and works as a circular queue where the queue head on each worker is rotated to a position corresponding to its id $n$.
When dealing with non-IID data, we apply \textit{data-injection} on worker $n$ by randomly pinging a subset of workers and pulling data, as determined by ($\alpha,\beta,\delta$) parameters.
To keep the batch-size from blowing up when $N$ is large, we reduce local mini-batch size to $b'$ of Eqn. (\ref{eqn:noniidinject}).
We implement data-injection via point-to-point (P2P) communication, i.e., \texttt{send}/\texttt{push} and \texttt{recv}/\texttt{pull} calls.

It is also worth pointing out that the synchronization steps currently implemented as PS strategy can be switched with an alternative communication operation.
For e.g., \texttt{pullFromPS} and \texttt{pushToPS} function calls in Alg. (\ref{alg:selsyncalgo}) can be easily swapped for an \texttt{AllReduce} collective.
The bandwidth cost of PS strategy increases with the number of workers, while allreduce collectives based on decentralized topology like ring or tree are bandwidth-optimal and logarithmic bandwidth-optimal respectively \cite{b46}.
By trading for more optimal communication strategies, training speedup can be further improved.
The main appeal with \texttt{SelSync}'s approach is its effectiveness in deeming which updates are important to communicate and which to use locally, thus switching between local and synchronous updates.

%% file: eval.tex
\section{Experimental Evaluation}\label{sec:eval}

\subsection{Cluster setup and workload specifications}\label{subsec:clustersetup}

We test \texttt{SelSync} on a system of 4 servers, each equipped with 48-core Intel Xeon E5-2560 CPU, 128GB system memory and 4 NVIDIA V100 GPUs.
To ensure resource isolation and minimum interference, we deploy a cluster of 16 CUDA-compatible nvidia-docker containers such that each container is allocated 8vCPUs, 24GB system memory and 1 NVIDIA V100 GPU.
The worker containers push/pull their updates to/from a PS container with 12vCPUs and 32GB system memory.
Communication between PS and worker containers is enabled via docker swarm configured over a 5Gbps NIC.

\textbf{Data for IID training}: CIFAR10/100 contains 50K train and 10K test images with 10 and 100 labels respectively.
ImageNet-1K has 1.28 million training and 50K test images spanning across 1K class labels.
WikiText-103 has 100 million tokens of verified and featured articles from Wikipedia \cite{b34}.

\textbf{Data for non-IID training}: To evaluate performance in non-IID settings, we split CIFAR10/100 across 10 workers with 1 and 10 labels per-worker for CIFAR10 and CIFAR100.

\textbf{DNNs and hyperparameters:}
We train ResNet101 on CIFAR10 and report \emph{top-1 test accuracy} from training with per-worker batch-size 32 and SGD optimizer with initial lr 0.1, weight decay 0.0004 and momentum 0.9.
We progressively decay the lr by 10$\times$ after 110 and 150 epochs.
VGG11 measures \emph{top-1 test accuracy} on CIFAR100 with batch-size 32 and SGD optimizer with momentum 0.9, weight decay 0.0005 and initial lr 0.01 that decays by 10$\times$ after epoch 50 and 75.
AlexNet trains on ImageNet-1K to report \emph{top-5 test accuracy} with batch-size 128, Adam optimizer \cite{b2} and fixed lr of 0.0001.
Lastly, we train Transformer on WikiText-103 where the encoder has 2 hidden layers of dimension 200, 2 heads, an embedding dimension of 200, dropout 0.2 and 35 bptt (backpropagation through time) steps.
We set per-worker batch-size to 20 and use SGD optimizer with initial lr 2.0 that decays by a factor of 0.8 every 2000 iterations and report the \emph{test perplexity}, i.e., exponential of the loss function.

\textbf{Comparing \texttt{SelSync} with SOTA learning algorithms:}
The same cluster of 16 workers and 1 PS is also used to evaluate BSP, FedAvg and SSP.
We train with BSP across all 16 workers using the default partitioning scheme (DefDP) and updates are aggregated on the PS at every iteration.
\texttt{SelSync} is deployed with different thresholds; a $\delta$ of 0.25, 0.3 or 0.5.
We launch \emph{FedAvg} with four different configurations by varying the fraction of participating workers (C) and the synchronization factor (E): ($C$, $E$) $\rightarrow$ (1, 0.25), (1, 0.125), (0.5, 0.25) and (0.5, 0.125).
This means we either use all or just half of the participating workers during communication phase in FedAvg, while aggregation is done either 4 or 8 times in every epoch at uniform intervals.
We deploy \emph{SSP} with two staleness thresholds: 100 and 200 steps.
The chosen $s$ values are large enough to not be a frequent synchronization barrier, while still small enough to not let worker models diverge too far from the global model.

\begin{figure}
	\subfloat[$\triangle (g_{i})$ computation overhead]{\includegraphics[width=0.25\textwidth]{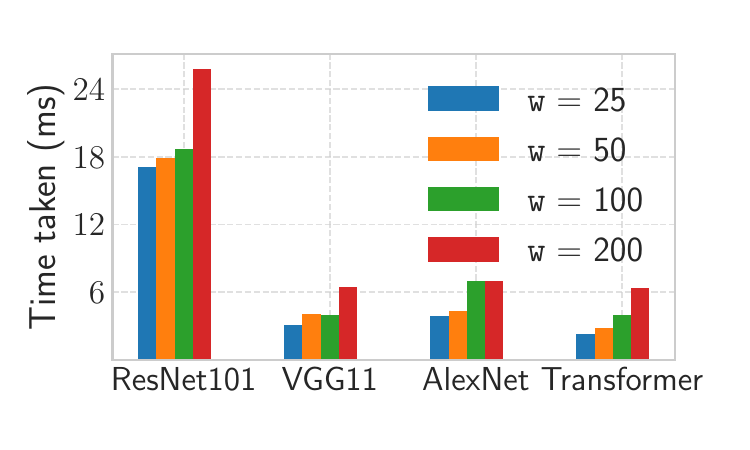}
	\label{fig:deltaoverhead}}
	\subfloat[Data-partitioning overhead]{\includegraphics[width=0.25\textwidth]{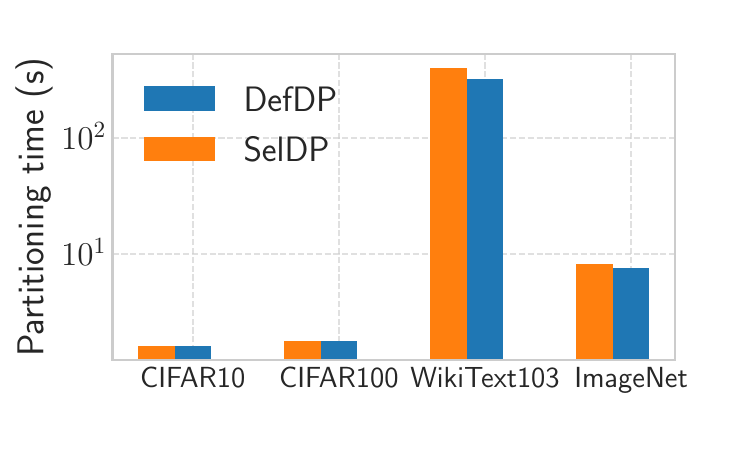}
	\label{fig:selsyncdpoverhead}}
	\caption{(a) Overhead of computing gradient variance and applying EWMA smoothing for different window-sizes. (b) Average time taken across workers to shuffle and create SelDP. 
	This is a one-time overhead incurred as part of data-batching and loading in the preprocessing stage.}
	\label{fig:selsyncoverhead}
\end{figure}

\subsection{Computational overhead of \texttt{SelSync}}\label{subsec:selsyoverhead}

The \texttt{RelativeGradChange} function from Alg. (\ref{alg:selsyncalgo}) applies smoothing because gradients computed over a mini-batch between consecutive iterations can be noisy.
This overhead is exclusive to \texttt{SelSync} as $\triangle(g_{i})$ is not computed in BSP, FedAvg or SSP.
As the window-size used for smoothing becomes larger, this overhead rises as well (shown in Fig. (\ref{fig:deltaoverhead})).
Time increases from 17 to 26ms in ResNet101 as the window-size increases from 25 to 200 steps; a jump of about 53\%.
A window of 25 iterations takes only about 3.1, 3.9 and 2.3ms on VGG11, AlexNet and Transformer respectively.
On increasing their windows to 200, their respective latencies increase by 110\%, 79\% and 178\%.
Although this cost increases with larger windows, this latency is considerably lower than the typical computation and communication cost of distributed training.
Besides, we observed in our evaluation that a $w$ = 25 sufficed for detecting inter-iteration gradient changes.
We use this window-size in our experiments, unless specified otherwise.

We evaluate the overhead of Selsync-data partitioner (SelDP) in this section as well.
We compare the overhead of the two partitioning schemes for \texttt{SelSync} in Fig. (\ref{fig:selsyncdpoverhead}).
Instead of splitting the data into unique, equal-sized chunks, SelDP rearranges the partitions based on worker id in the cluster.
For CIFAR10/100 datasets, this additional operation does \textit{not} incur considerable overhead as loading time of SelDP and DefDP are nearly identical.
This is because size of these datasets is not that big to pose a considerable overhead to reorder and shuffle.
For bigger datasets like ImageNet-1K and WikiText-103, SelDP had a higher processing time over the default partitioning scheme.
However, the margin is only of a few seconds and the partitioning operation is a one-time overhead incurred before the actual training begins.

\subsection{Training \texttt{SelSync} with SelDP vs. DefDP}\label{subsec:seldpVSdefdp}

\begin{figure}
	\subfloat[ResNet101]{\includegraphics[width=0.23\textwidth]{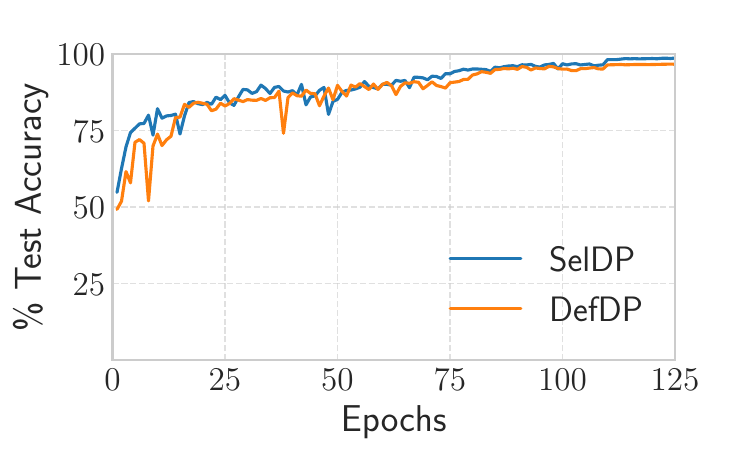}
	\label{resnetSPDP}}
	\hfill
	\subfloat[VGG11]{\includegraphics[width=0.23\textwidth]{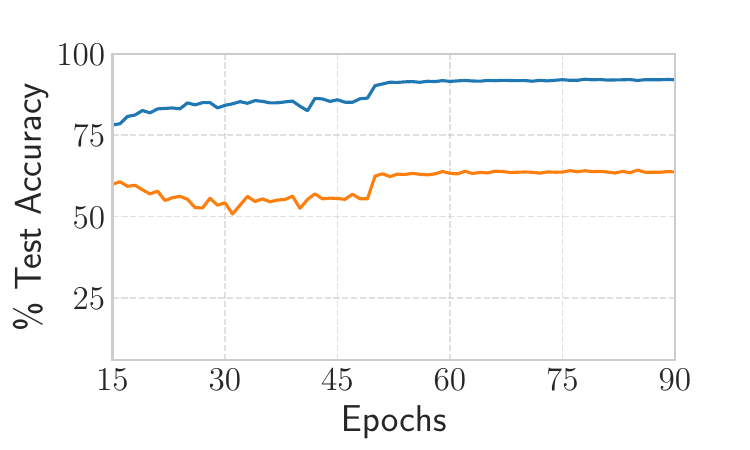}
	\label{vggSPDP}}
	\hspace{0.1cm}
	\subfloat[AlexNet]{\includegraphics[width=0.23\textwidth]{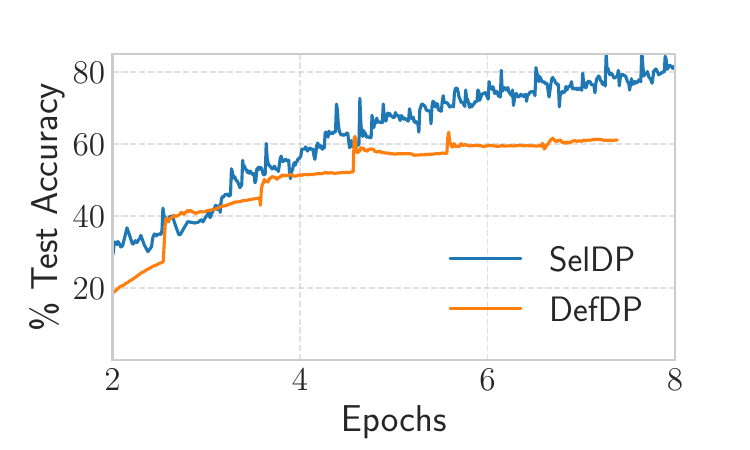}
	\label{alexnetSPDP}}
	\hfill
	\subfloat[Transformer]{\includegraphics[width=0.23\textwidth]{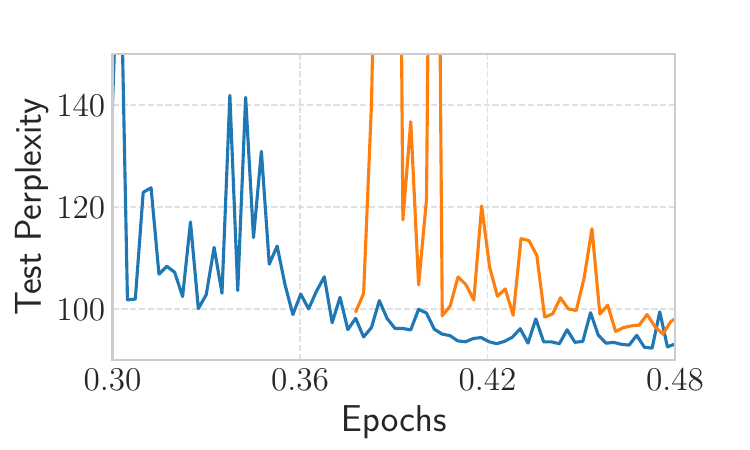}
	\label{transformerSPDP}}
	\caption{Test accuracy/perplexity curves of \texttt{SelSync} for $\delta$ = 0.25 with SelDP and DefDP.
	We perform \emph{gradient} aggregation during synchronization phase.
	For the same epochs, SelDP achieves better generalization/test performance.}
	\label{fig:spvsdp1}
\end{figure}

In this section, we deploy \texttt{SelSync} with $\delta$ = 0.25 and see how different models converge with our custom partitioning vs. the default scheme typically used in DDP training.
We plot the convergence curves in Fig. (\ref{fig:spvsdp1}) (i.e., test accuracy or perplexity vs. epochs).
\textit{During the synchronization phase, we perform gradient aggregation on the PS and return reduced gradients back to the workers to apply locally}.
By the end of epoch 125, ResNet101 converged to 97.6\% test accuracy on SelDP and 96.8\% on the default scheme.
CIFAR100 partitioned with DefDP performed poorly on VGG11 with only 64.1\% accuracy, while SelDP converged to 90.86\% after 90 epochs.
We see more performance degradation in VGG11 with DefDP compared to ResNet101 as the training data of the former has more labels to evaluate; CIFAR100 vs. CIFAR10.
Although VGG11 is larger, ResNet101 is a skip-connection based architecture that generalizes better compared to the simpler convolution-based architecture \cite{b10}.
With SelDP partitioning, AlexNet converged to 81.1\% top-5 test accuracy while \texttt{SelSync} with DefDP did not improve beyond 61.2\%.
Training Transformer on \texttt{SelSync} with SelDP attained 92.6 while DefDP converged to 94.91 test perplexity near the end of half an epoch.
The perplexity before epochs 0.3 and 0.38 for the two partitioning schemes is omitted in Transformer as it was very high initially and fell rapidly in that phase.

\textbf{Summary}: \textit{SelDP outperforms DefDP in semi-synchronous training.
This is because many updates are performed locally and not aggregated among workers.
With DefDP, worker replicas only train on data local to them and fail to learn from partitions located on other workers.
As a result, local models tend to diverge away from the global model.
With SelDP, each worker has access to the entire training data, albeit in a different order.
This ensures the local model weights do not significantly diverge in their local minima.}

 \begin{figure}
	\subfloat[ResNet101]{\includegraphics[width=0.23\textwidth]{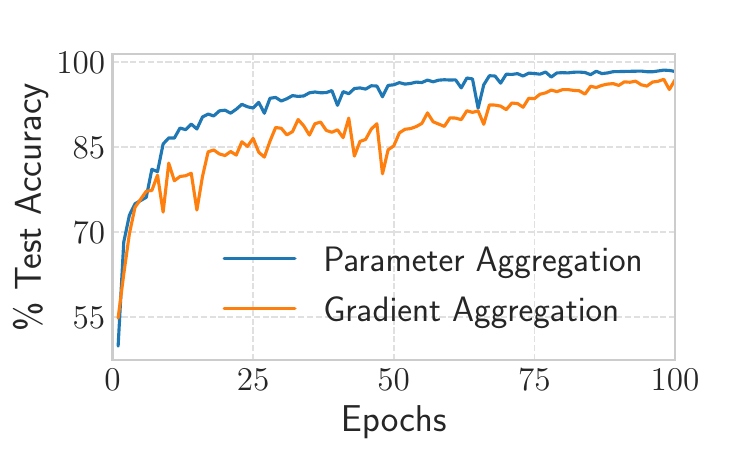}
	\label{resnetmodelgradavg}}
	\hfill
	\subfloat[VGG11]{\includegraphics[width=0.23\textwidth]{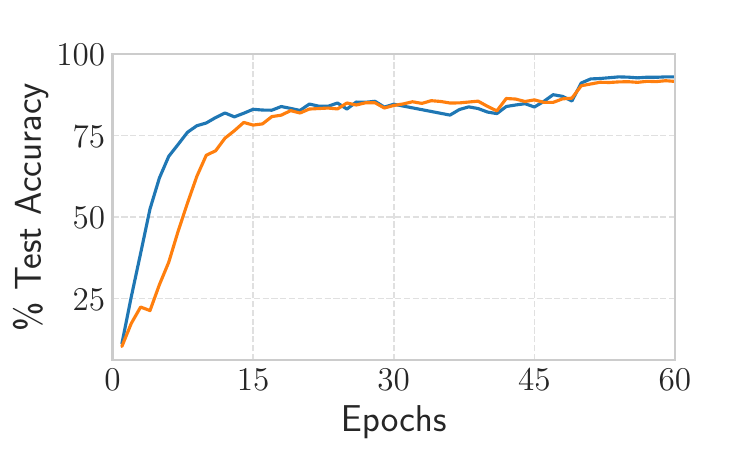}
	\label{vggmodelgradavg}}
	\hspace{0.1cm}
	\subfloat[AlexNet]{\includegraphics[width=0.23\textwidth]{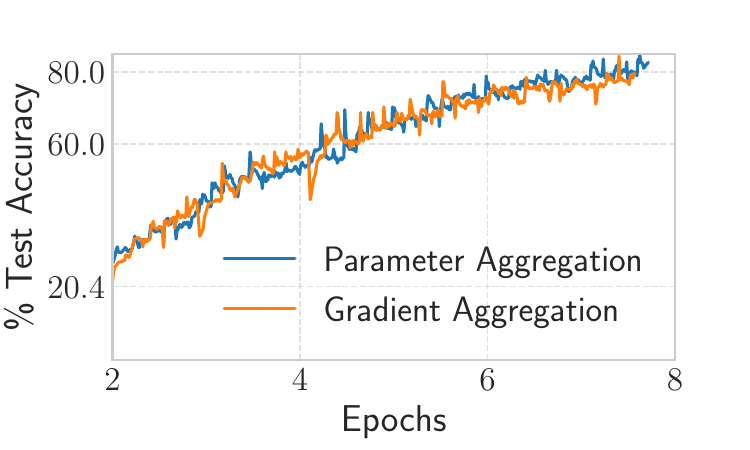}
	\label{alexnetmodelgradavg}}
	\hfill
	\subfloat[Transformer]{\includegraphics[width=0.23\textwidth]{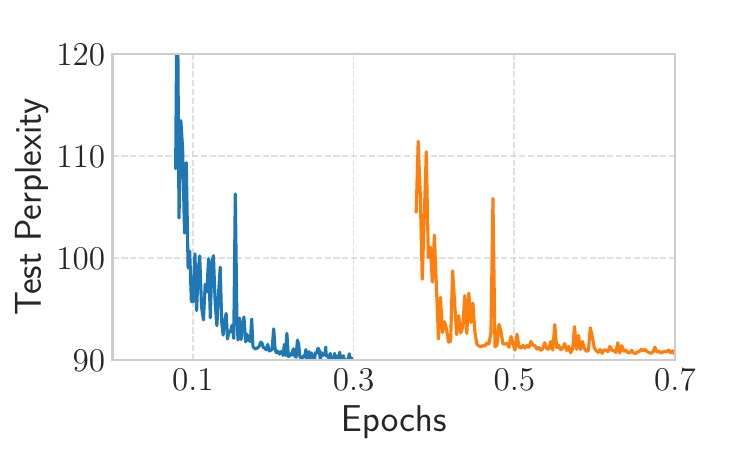}
	\label{transformermodelgradavg}}
	\caption{Convergence curves comparing gradients (GA) vs. parameter aggregation (PA) in \texttt{SelSync} with $\delta$ = 0.25.
	For the same epochs, PA achieved better accuracy in settings with a learning rate decay schedule.
	Interestingly, both performed similarly in AlexNet where the learning rate was fixed.}
	\label{fig:gavspa}
\end{figure}

\subsection{\texttt{SelSync} with Gradient vs. Parameter Aggregation}\label{subsec:gradparamagg}

We fix every other setting and deploy \texttt{SelSync} where we only change the aggregation mechanism: \textit{gradient aggregation (GA)} vs. \textit{parameter aggregation (PA)}.
We set $\delta$ = 0.25, partition training data with SelDP and compare their accuracy/perplexity over a fixed set of epochs.
The convergence plots are shown in Fig. (\ref{fig:gavspa}).
By the end of 100 epochs, ResNet101 converges to 96.8\% test accuracy with GA and 98.52\% with PA, an improvement of 1.72\% with parameter averaging.
VGG11 with GA attained 90.86\% accuracy and 91.42\% with PA at epoch 60, about 0.56\% more than GA for the same amount of training.
We see in Fig. (\ref{transformermodelgradavg}) that both GA and PA converge to a perplexity of 90.0 for Transformer; however, \texttt{SelSync} with PA reached there in 0.3 epochs while GA took 130\% more iterations to converge.
AlexNet, the only model trained with a fixed learning rate, had similar convergence trajectory on both GA and PA.

\begin{figure}
	\subfloat[Epoch 25]{\includegraphics[width=0.23\textwidth]{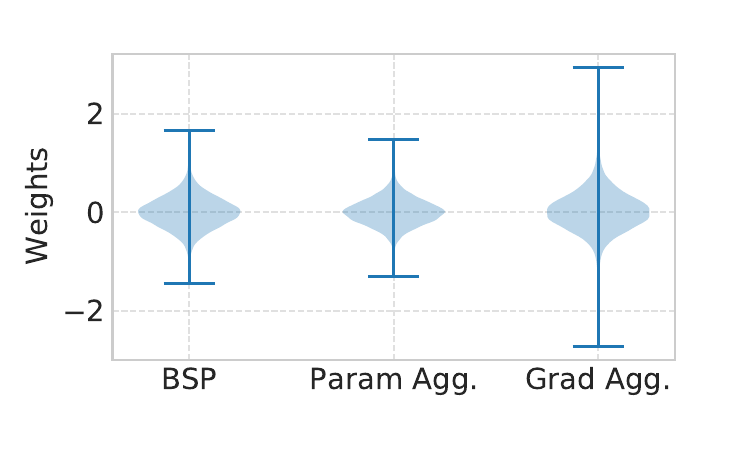}
	\label{epoch25weights}}
	\hfill
	\subfloat[Epoch 50]{\includegraphics[width=0.23\textwidth]{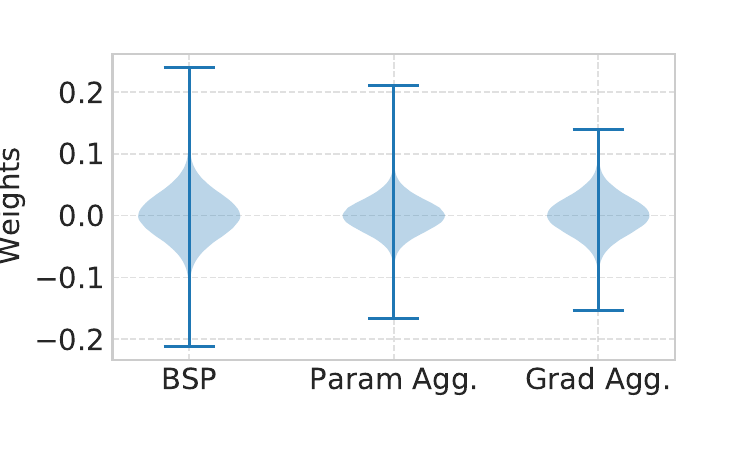}
	\label{epoch50weights}}
	\caption{Comparing distribution density of model weights while training independently via BSP, \texttt{SelSync} with Param Agg. (PA) and Grad Agg. (GA).
	Distribution between model weights is compared over epochs 25 and 50 for layer \texttt{layer1\_1\_conv1\_weight} of ResNet101.
	BSP and \texttt{SelSync} with PA have similar distributions compared to \texttt{SelSync} with GA.}
	\label{fig:weightdist}
\end{figure}

The better generalization of \texttt{SelSync} with PA over GA can be seen from the distribution of model parameters of the two approaches and how well they align with the parameter distributions had we trained with BSP.
For three independent runs of ResNet101 with BSP, \texttt{SelSync} with GA and PA, Fig. (\ref{fig:weightdist}) shows the parameter density estimates plotted at epochs 25 and 50.
On 25th epoch, BSP and PA have similar distributions, whereas parameters of GA were comparatively spread apart.
As training continues (around epoch 50), parameter distribution of GA is much smaller and narrower than both BSP and \texttt{SelSync} with PA.
Thus, we see that from the distributions that weights diverge more with gradient aggregation.
Comparatively, parameter aggregation was much more aligned with BSP training.
This is because by aggregating parameters during synchronization phase, we bound the divergence between local models on the workers and the global model on the PS.

\subsection{Training with IID and non-IID data}\label{subsec:selsyncperfiid}

\begin{table*}
	\centering
	\begin{tabular}{|c|c|c|c|c|c|c|c|}
	\hline
	\bfseries Model & \bfseries Method & \bfseries Iterations &  \bfseries LSSR & \bfseries Acc./PPL & \bfseries Conv. Diff. & \bfseries Outperform BSP ? & \bfseries Overall speedup \\
	\hline
	\multirow{9}{*}{ResNet101} & BSP & 22.8K & 0.0 & 98.5\% & 0.0\% & N/A & 1$\times$ \\
	\cline{2-8}
	& FedAvg (1, 0.25) & 63.4K & 0.979 & 98.55\% & +0.05\% & True & 2.59$\times$ \\
	\cline{2-8}
	& FedAvg (1, 0.125) & 57.1K & 0.959 & 98.6\% & +0.1\% & True & 2.55$\times$ \\
	\cline{2-8}
	& FedAvg (0.5, 0.25) & 66.8K & 0.979 & 94.4\% & -4.1\% & False & - \\
	\cline{2-8}
	& FedAvg (0.5, 0.125) & 62.8.1K & 0.959 & 95.5\% & -3.0\% & False & - \\
	\cline{2-8}
	& SSP $s$ = 100 & 54.8K & - & 93.8\% & -4.7\% & False & - \\
	\cline{2-8}
	& SSP $s$ = 200 & 59.4K & - & 93.33\% & -5.17\% & False & - \\
	\cline{2-8}
	& \textbf{\texttt{SelSync} $\delta$ = 0.3} & \textbf{42.2K} & \textbf{0.829} & \textbf{98.62\%} & \textbf{+0.12\%} & \textbf{True} & \textbf{2.03$\times$} \\
	\cline{2-8}
	& \textbf{\texttt{SelSync} $\delta$ = 0.5} & \textbf{53.2K} & \textbf{0.846} & \textbf{98.56\%} & \textbf{+0.06\%} & \textbf{True} & \textbf{1.67$\times$} \\
	\hline
	\multirow{9}{*}{VGG11} & BSP & 10.8K & 0.0 & 90.9\% & 0.0\% & N/A & 1$\times$ \\
	\cline{2-8}
	& FedAvg (1, 0.25) & 17.4K & 0.979 & 87.56\% & -3.34\% & False & - \\
	\cline{2-8}
	& FedAvg (1, 0.125) & 20.2K & 0.959 & 88.2\% & -2.7\% & False & - \\
	\cline{2-8}
	& FedAvg (0.5, 0.25) & 19.6K & 0.979 & 84.6\% & -6.3\% & False & - \\
	\cline{2-8}
	& FedAvg (0.5, 0.125) & 21.1K & 0.959 & 85.16\% & -5.74\% & False & - \\
	\cline{2-8}
	& SSP $s$ = 100 & 11.8K & - & 68.72\% & -22.18\% & False & - \\
	\cline{2-8}
	& SSP $s$ = 200 & 12.2K & - & 68.78\% & -22.12\% & False & - \\
	\cline{2-8}
	& \textbf{\texttt{SelSync} $\delta$ = 0.3} & \textbf{72.4K} & \textbf{0.912} & \textbf{91.33\%} & \textbf{+0.43\%} & \textbf{True} & \textbf{6.16$\times$} \\
	\cline{2-8}
	& \textbf{\texttt{SelSync} $\delta$ = 0.5} & \textbf{81.2K} & \textbf{0.937} & \textbf{91.32\%} & \textbf{+0.42\%} & \textbf{True} & \textbf{13.75$\times$} \\
	\hline
	\multirow{9}{*}{AlexNet} & BSP & 106.5K & 0.0 & 85.15\% & 0.0\% & N/A & 1$\times$ \\
	\cline{2-8}
	& FedAvg (1, 0.25) & 118K & 0.997 & 77.15\% & -8.0\% & False & - \\
	\cline{2-8}
	& FedAvg (1, 0.125) & 124K & 0.993 & 78.2\% & -6.95\% & False & - \\
	\cline{2-8}
	& FedAvg (0.5, 0.25) & 115.6K & 0.997 & 75.6\% & -9.55\% & False & - \\
	\cline{2-8}
	& FedAvg (0.5, 0.125) & 121.4K & 0.993 & 76.67\% & -8.48\% & False & - \\
	\cline{2-8}
	& SSP $s$ = 100 & 111.5K & - & 86.52\% & +1.37\% & True & 4.53$\times$ \\
	\cline{2-8}
	& SSP $s$ = 200 & 113.2K & - & 85.8\% & +0.65\% & True & 4.47$\times$ \\
	\cline{2-8}
	& \textbf{\texttt{SelSync} $\delta$ = 0.3} & \textbf{95.8K} & \textbf{0.954} & 86.72\% & \textbf{+1.57\%} & \textbf{True} & \textbf{4.62$\times$} \\
	\cline{2-8}
	& \textbf{\texttt{SelSync} $\delta$ = 0.5} & \textbf{106.8K} & \textbf{0.966} & \textbf{85.93\%} & \textbf{+0.78\%} & \textbf{True} & \textbf{4.63$\times$} \\
	\hline
	\multirow{9}{*}{Transformer} & BSP & 28K & 0.0 & 90.02 & 0.0\% & N/A & 1$\times$ \\
	\cline{2-8}
	& FedAvg (1, 0.25) & 112.1K & 0.998 & 89.96 & +0.06 & True & 1.99$\times$ \\
	\cline{2-8}
	& FedAvg (1, 0.125) & 104.2K & 0.999 & 90.01 & +0.01 & True & 2.14$\times$ \\
	\cline{2-8}
	& FedAvg (0.5, 0.25) & 118K & 0.999 & 91.67 & -1.65 & False & - \\
	\cline{2-8}
	& FedAvg (0.5, 0.125) & 111.6K & 0.998 & 91.2 & -1.18 & False & - \\
	\cline{2-8}
	& SSP $s$ = 100 & 110K & - & 89.98 & +0.04 & True & 2.02$\times$ \\
	\cline{2-8}
	& SSP $s$ = 200 & 116.2K & - & 90.0 & +0.0 & True & 1.92$\times$ \\
	\cline{2-8}
	& \textbf{\texttt{SelSync} $\delta$ = 0.3} & \textbf{38.6K} & \textbf{0.725} & \textbf{89.91} & \textbf{+0.11} & \textbf{True} & \textbf{2.42$\times$} \\
	\cline{2-8}
	& \textbf{\texttt{SelSync} $\delta$ = 0.5} & \textbf{46.1K} & \textbf{0.733} & \textbf{89.97} & \textbf{+0.05} & \textbf{True} & \textbf{2.06$\times$} \\
	\hline
	\end{tabular}
	\vspace{0.15cm}
	\caption{DNN Performance Across \texttt{SelSync}, BSP, FedAvg and SSP.
	Convergence Difference (Conv. Diff.) and Speedup w.r.t. BSP.}
	\label{table:ddpconvergence}
\end{table*}

\textbf{IID training}: To measure the fraction of training using local updates versus steps that are synchronized among workers, we define a metric called \textbf{local-to-synchronous step ratio}, or \textbf{LSSR} measured as:

\begin{equation}
	\text{LSSR} = \frac{\texttt{steps}_{local}}{\texttt{steps}_{local} + \texttt{steps}_{bsp}}
	\label{eqn:lsratio}
\end{equation}

If all updates are local, then $\mathtt{steps}_{bsp}$=0 and LSSR=1.
For BSP training, LSSR = 0 as all the updates are synchronous and there are \emph{no local updates}.
LSSR essentially measures communication reduction w.r.t. BSP as $1 / (1 - \text{LSSR})$ for the same number of iterations/epochs.
For e.g., LSSR of 0.9 implies a communication reduction of 10$\times$ over BSP.
\emph{LSSR scores \textbf{do not} apply to SSP as workers push updates to the PS asynchronously and do not synchronize them.}
When staleness threshold is crossed, faster workers wait for slower workers to catch-up. There is no explicit gradient summation/averaging in SSP, thus a metric like LSSR doesn’t apply.
Table (\ref{table:ddpconvergence}) compares \texttt{SelSync} with BSP, FedAvg and SSP.
We run each model until the accuracy/perplexity does not improve any further and note the corresponding iterations taken and the LSSR scores (if applicable).
We measure convergence and speedup w.r.t. BSP as many semi-synchronous methods do provide training speedups, but not necessarily meet the same convergence targets as BSP.
Results from the \textit{`Overall speedup'} column of Table (\ref{table:ddpconvergence}) are omitted for settings that failed to converge to the same accuracy as BSP.

For ResNet101, we see that \texttt{SelSync} and few configurations of FedAvg achieve higher accuracy than BSP.
The improvement over BSP comes from abundant exploration of the local minima on individual workers which results in better generalization performance \cite{b33}.
However, too much local exploration as well as infrequent (from $E$) and partial (from $C$) aggregation can increase the divergence between the local and global model state, as seen in FedAvg for ($C,E$) values (0.5,0.25) and (0.5,0.125).
Here, local updates were collected from only half the workers during synchronization.
FedAvg with $C$=1 performed well even with high LSSR scores ($\approx$ 0.95-0.97) as ResNet101 is an over-parameterized network for a dataset like CIFAR10, and was thus more robust to local training with FedAvg.
\texttt{SelSync} with $\delta$ = 0.3 and 0.5 had LSSR scores 0.83-0.84.
Thus, overall speedup over BSP was slightly higher for FedAvg on account of its higher LSSR scores; around 2.59$\times$ while \texttt{SelSync} was up to 2$\times$.
SSP failed to beat BSP because asynchronously pulling/pushing updates from/to the PS on a layer-by-layer basis introduced considerable staleness while training ResNet101.

A simpler convolutional network like VGG11 failed to achieve BSP-level accuracy with both FedAvg and SSP.
\texttt{SelSync} attained 0.43\% more accuracy over BSP and reduced training time by up to 13.75$\times$ due to high LSSR scores ($\approx$ 0.91-0.93).
They are however lower than the LSSR scores of FedAvg because \texttt{SelSync} intelligently switches between synchronous and local updates based on their significance.
This is also why \texttt{SelSync} gets better test accuracy than BSP while FedAvg and SSP failed to do so.

AlexNet on ImageNet-1K failed to converge across all FedAvg configurations.
Since we set $E$ as the aggregation frequency in an epoch, the number of synchronization calls made with FedAvg were too low on account of large dataset size of 1.28 million samples.
This can also be seen from the extremely high LSSR scores ($\approx$ 0.99) and thus, model trained locally for almost all iterations and learned features  \textit{only} from its local data.
Due to smaller size and fewer layers of AlexNet, staleness was not an issue with SSP-based training with either of the $s$-thresholds.
By asynchronously pushing and pulling updates to/from the PS, workers were also able to learn features from partitions on other workers, thus further improving the test accuracy.
SSP attained up to 1.37\% better accuracy than BSP while speeding up training by 4.53$\times$.
\texttt{SelSync} was 4.63$\times$ faster and had 1.57\% better accuracy than BSP.
The speedup came from synchronizing only the crucial updates and performing 95-96\% updates locally (as can be seen from the LSSR scores).

Last, Transformer with $C$ = 1 in FedAvg converged to a better perplexity than BSP while being 2.14$\times$ faster, courtesy of its high LSSR scores.
SSP also successfully beat BSP, and converged 2$\times$ faster due to its asynchronous execution.
\texttt{SelSync} trained for roughly 39-46K steps and synchronized about 26\% of those iterations.
By doing so, it converged to even lower perplexity than BSP while accelerating training by up to 2.42$\times$.

\begin{figure}
	\subfloat[ResNet101 on CIFAR10]{\includegraphics[width=0.23\textwidth]{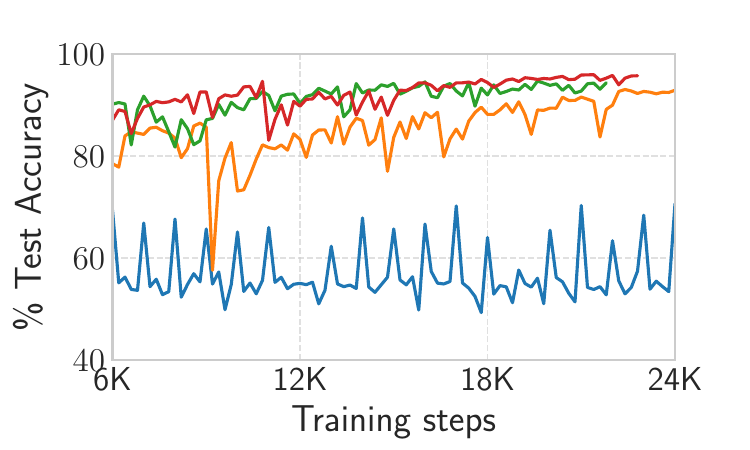}
		\label{fig:resnetnoniid}}
	\hfill
	\subfloat[VGG11 on CIFAR100]{\includegraphics[width=0.23\textwidth]{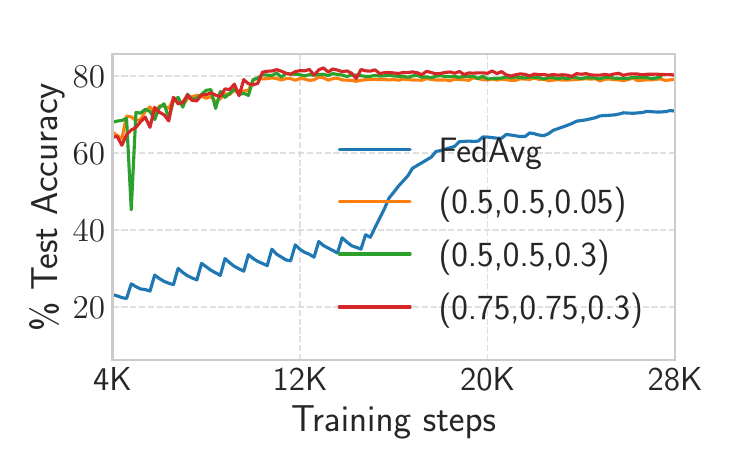}
	\label{fig:vggnoniid}}
	\caption{Data-injection in \texttt{SelSync} for different ($\alpha, \beta, \delta$) configurations vs. FedAvg on non-IID data.}
	\label{fig:noniideval}
\end{figure}

\textbf{Non-IID training}: We now compare convergence between FedAvg and \texttt{SelSync} in non-IID settings.
We choose three data-injection configurations ($\alpha,\beta,\delta$) to execute \texttt{SelSync}: (0.5, 0.5, 0.05), (0.5, 0.5, 0.3) and (0.75, 0.75, 0.3).
As per Eqn. (\ref{eqn:noniidinject}), we set $b'$ = 11 to keep $b$ = 32 for the first two configurations and $b'$ = 6 for the last.
With non-IID data, FedAvg oscillated between 60-70\% test accuracy for ResNet101, and saturated to 70\% accuracy on VGG11.
As ($\alpha,\beta,\delta$) increases, the corresponding data distribution improves and thus, so does the accuracy.
\texttt{SelSync} configuration (0.75, 0.75, 0.3) thus achieved the maximum accuracy, followed by (0.5, 0.5, 0.3) and (0.5, 0.5, 0.05) in respective order.
We thu see that Data-injection works well specifically for \texttt{SelSync} and generally in the context of semi-synchronous training.

%% file: conclusion.tex
\section{Conclusion}\label{sec:conclusion}

From Table (\ref{table:ddpconvergence}), we see that BSP takes least number of steps to converge since it performs maximum work per-step by combining updates from all workers.
However, it also suffers from high communication overheads to collect and distribute back updates.
On the other hand, FedAvg provides decent speedups due to its high LSSR scores, however, the ($C,E$) configuration that provides maximum speedup or even convergence guarantees varies for each DNN and dataset.
We saw in our evaluation that naively aggregating updates after fixed steps/epochs does not always yield the same convergence as BSP.
Although workers communicate with the PS asynchronously in SSP, pushing/pulling updates for large and deeper models to/from the PS introduces staleness issues that either slows-down or even saturates convergence.
Both FedAvg and SSP take more steps than BSP as they perform lesser work per-training step (i.e., $b$ vs. $Nb$ samples processed on each step).
Across all models, \texttt{SelSync} achieved better accuracy/perplexity than BSP by selectively synchronizing updates that were deemed critical, applied updates locally otherwise.
The relative gradient change metric $\triangle (g_{i})$ thus serves as an effective indicator of measuring the significance of each update in DNN training.
With its high LSSR scores, \texttt{SelSync} eliminates considerable communication cost in the training phase and allows workers to explore local minima more than BSP, while still bounding the divergence between local and global model through frequent synchronization.